\documentclass{article}
\usepackage{arxiv}
\usepackage[utf8]{inputenc}
\usepackage[T1]{fontenc}
\usepackage{hyperref}
\usepackage{url}
\usepackage{booktabs}
\usepackage{amsfonts}
\usepackage{amsmath}
\usepackage{amssymb}
\usepackage{nicefrac}
\usepackage{microtype}
\usepackage{cleveref}
\usepackage{graphicx}
\usepackage{natbib}
\usepackage{doi}
\usepackage{bm}
\usepackage{subcaption}
\usepackage{epstopdf}

\graphicspath{{figures/}}

\title{$\mathcal{H}_\infty$ Robust Control for Gust Load Alleviation of Geometrically Nonlinear Flexible Aircraft}

\author{
  Nikolaos D.~Tantaroudas\thanks{Corresponding author. Senior Researcher, ICCS.} \\
  Institute of Communications and Computer Systems (ICCS)\\
  9 Iroon Politechniou Street, Zografou, Athens 15773, Greece \\
  \texttt{nikolaos.tantaroudas@iccs.gr} \\
  \And
  Ilias Karachalios \\
  National Technical University of Athens\\
  Zografou, Athens 15780, Greece \\
}

\renewcommand{\shorttitle}{$\mathcal{H}_\infty$ Control for Gust Load Alleviation}

\hypersetup{
  pdftitle={H-infinity Robust Control for Gust Load Alleviation of Geometrically Nonlinear Flexible Aircraft},
  pdfsubject={eess.SY, physics.flu-dyn},
  pdfauthor={N.D. Tantaroudas, I. Karachalios},
  pdfkeywords={H-infinity control, Gust load alleviation, Flexible aircraft, Robust control, Reduced-order model}
}

\begin{document}
\maketitle

\begin{abstract}
$\mathcal{H}_\infty$ robust control synthesis for gust load alleviation (GLA) of very flexible aircraft is presented. The controller is synthesised on a compact reduced-order model comprising 8 degrees of freedom for the UAV configuration and 9 for the flying-wing, obtained through nonlinear model order reduction of the coupled fluid-structure-flight dynamics system, and validated on the full nonlinear model (540 and 1,616 degrees of freedom, respectively). The control architecture employs trailing-edge flap deflection as the actuator and wing-tip displacement as the performance output, with an input-shaping weighting function $K_c$ that governs the trade-off between structural load alleviation and rigid-body trajectory deviation. Results are presented for a Global Hawk-like UAV (23.15\% wing-tip deflection reduction under discrete gust, 10.26\% under stochastic turbulence) and a very flexible flying-wing configuration. The methodology demonstrates that $\mathcal{H}_\infty$ controllers designed on low-order ROMs can robustly alleviate gust loads when applied to high-dimensional nonlinear aeroelastic systems.
\end{abstract}


\section*{Nomenclature}
\label{sec:nomenclature}

\noindent\textit{Latin}

\begin{tabbing}
12345678901234 \= \kill
$b$            \> = semichord \\
$\mathbf{A}$   \> = Jacobian matrix \\
$\mathbf{B}_c,\,\mathbf{B}_g$ \> = control and gust input matrices \\
$D_{kij}$      \> = bilinear interaction coefficients of the ROM \\
$H_g$          \> = gust gradient distance \\
$K_c$          \> = input-shaping weighting parameter \\
$K(s)$         \> = $\mathcal{H}_\infty$ controller transfer function \\
$L_w$          \> = turbulence scale length \\
$m$            \> = number of retained ROM modes \\
$n$            \> = number of full-order degrees of freedom \\
$\bm{R}$       \> = residual vector \\
$t$            \> = physical time \\
$u$            \> = control input (flap angular acceleration, $\ddot{\delta}$) \\
$U_\infty$     \> = freestream velocity \\
$\bm{w}$       \> = full-order state vector \\
$w_0$          \> = peak (design) gust velocity \\
$w_g$          \> = gust vertical velocity \\
$\mathbf{x}$   \> = augmented state vector \\
$\mathbf{z}$   \> = reduced-order state vector \\
$\mathbf{z}_{\text{perf}}$ \> = performance output vector
\end{tabbing}

\noindent\textit{Greek}

\begin{tabbing}
12345678901234 \= \kill
$\alpha$       \> = angle of attack \\
$\alpha_0$     \> = trim angle of attack \\
$\gamma^*$     \> = optimal $\mathcal{H}_\infty$ norm \\
$\delta$       \> = trailing-edge flap deflection \\
$\lambda_k$    \> = $k$-th eigenvalue of the Jacobian \\
$\boldsymbol{\Lambda}$ \> = diagonal matrix of retained eigenvalues \\
$\boldsymbol{\Phi},\,\boldsymbol{\Psi}$ \> = right and left eigenvector matrices \\
$\rho$         \> = air density \\
$\sigma_w$     \> = turbulence intensity (RMS gust velocity) \\
$\bar{\sigma}$ \> = maximum singular value
\end{tabbing}

\noindent\textit{Acronyms}

\begin{tabbing}
12345678901234 \= \kill
DOF  \> = degree(s) of freedom \\
GLA  \> = gust load alleviation \\
HALE \> = high-altitude long-endurance \\
MAC  \> = mean aerodynamic chord \\
NMOR \> = nonlinear model order reduction \\
ROM  \> = reduced-order model \\
UAV  \> = unmanned aerial vehicle \\
VFA  \> = very flexible aircraft
\end{tabbing}

\section{Introduction}
\label{sec:intro}

Gust load alleviation (GLA) is essential for the structural integrity, fatigue life, and ride quality of next-generation very flexible aircraft. These platforms, including high-altitude long-endurance (HALE) UAVs, solar-powered aircraft, and high-aspect-ratio flying wings, undergo large structural deformations that produce strong coupling between structural flexibility, unsteady aerodynamics, and flight dynamics~\citep{Patil2001, Noll2004, Patil2006}. Without active control intervention, atmospheric gusts can induce large wing-tip deflections, leading to excessive structural loads and potential aeroelastic instabilities~\citep{Livne2018}. The catastrophic failure of the NASA Helios prototype during a turbulence encounter~\citep{Noll2004} underscored the critical need for effective GLA systems on such platforms.

The challenge of GLA control design for very flexible aircraft lies at the intersection of high-dimensional modelling and nonlinear dynamics. Full-order coupled models with hundreds to thousands of degrees of freedom are computationally intractable for standard robust control synthesis algorithms, which scale poorly with system dimension. Reduced-order models (ROMs) provide a practical bridge: they compress the essential dynamics into a compact state-space that is amenable to control synthesis while preserving the key physical coupling mechanisms~\citep{DaRonch2013control, DaRonch2012rom, Tantaroudas2017bookchapter}.

$\mathcal{H}_\infty$ synthesis offers a natural framework for GLA because it explicitly addresses the worst-case disturbance rejection problem: the controller minimises the peak gain from gust input to structural response output, providing guaranteed robustness margins against modelling uncertainties. Previous applications of $\mathcal{H}_\infty$ control to aeroelastic systems include flutter suppression~\citep{DaRonch2014flutter, Papatheou2013ifasd} and manoeuvre load control~\citep{DaRonch2014scitech_flight}. The nonlinear behaviour of aerofoil systems, including limit-cycle oscillations, was investigated experimentally and numerically in~\citep{Fichera2014isma}.

Since 2015, notable advances have been made in control methodologies for flexible aircraft. Wang et al.~\citep{Wang2016aiaa} developed a nonlinear modal aeroservoelastic framework for flexible aircraft, subsequently applied to nonlinear aeroelastic control using model updating~\citep{Wang2018joa}. Waitman and Marcos~\citep{Waitman2020} demonstrated structured $\mathcal{H}_\infty$ synthesis for active flutter suppression of a flexible-wing UAV demonstrator, confirming the practical applicability of $\mathcal{H}_\infty$ methods to flexible platforms. Artola et al.~\citep{Artola2021} showed that minimal nonlinear modal descriptions are sufficient for aeroelastic control and estimation via moving-horizon estimation and model-predictive control. A comprehensive treatment of the coupled flight mechanics, aeroelasticity, and control of flexible aircraft is provided in the monograph by Palacios and Cesnik~\citep{PalaciosCesnik2023}.

This paper presents a systematic $\mathcal{H}_\infty$ GLA design framework based on the NMOR methodology developed in~\citep{DaRonch2013gust, DaRonch2013control} and extended to free-flying aircraft in~\citep{Tantaroudas2015scitech, Tantaroudas2014aviation}. The complete aeroservoelastic analysis framework is documented in~\citep{Tantaroudas2017bookchapter}. A self-contained derivation of the NMOR formulation, including third-order Taylor expansion terms and eigenvector selection criteria, is presented in the companion paper~\citep{Tantaroudas2026nmor}, while the application of the same ROM framework to rapid worst-case gust identification is demonstrated in~\citep{Tantaroudas2026gust}. The controller is designed on the linear ROM and applied to the full nonlinear model. Two test cases are considered: a Global Hawk-like UAV and a very flexible flying-wing. The influence of the input-shaping weighting function on the performance-trajectory trade-off is investigated in detail.

\section{Aeroelastic Model and Reduction}
\label{sec:model}

\subsection{Coupled System}

The aeroelastic-flight dynamic model integrates three subsystems. The structural dynamics employ geometrically-exact nonlinear beam theory~\citep{Hodges2003, Palacios2010} with displacement-based finite elements carrying six DOF per node. The aerodynamics use unsteady strip theory based on Theodorsen's framework~\citep{Theodorsen1935}, with Wagner~\citep{Wagner1925} and K\"ussner~\citep{Kussner1936} indicial functions providing time-domain unsteady loads. The flight dynamics use Newton--Euler equations with quaternion-based attitude propagation~\citep{Tantaroudas2015scitech}.

The coupled system is expressed in first-order state-space form:
\begin{equation}
\dot{\mathbf{w}} = \mathbf{R}(\mathbf{w}, \mathbf{u}_c, \mathbf{u}_d)
\label{eq:nonlinear_system}
\end{equation}
where $\mathbf{w} \in \mathbb{R}^n$ is the state vector partitioned into aerodynamic, structural, and rigid-body components; $\mathbf{u}_c$ is the control input (flap deflection); and $\mathbf{u}_d$ is the gust disturbance.

Linearisation about the trimmed equilibrium $\mathbf{w}_0$ yields:
\begin{equation}
\dot{\mathbf{w}} = \mathbf{A}\mathbf{w} + \mathbf{B}_c \mathbf{u}_c + \mathbf{B}_g \mathbf{u}_d + \mathbf{F}_{\text{nl}}(\mathbf{w})
\label{eq:linearised}
\end{equation}
where $\mathbf{A}$ is the Jacobian, $\mathbf{B}_c$ and $\mathbf{B}_g$ are input matrices, and $\mathbf{F}_{\text{nl}}$ contains the higher-order nonlinear terms.

\subsection{Model Order Reduction}

The NMOR procedure~\citep{DaRonch2013control, DaRonch2013gust, Tantaroudas2015scitech} projects the system onto $m$ biorthonormal eigenvectors of the Jacobian: $\Delta\mathbf{w} \approx \boldsymbol{\Phi}\mathbf{z}$, $\boldsymbol{\Psi}^H\boldsymbol{\Phi} = \mathbf{I}_m$, yielding:
\begin{equation}
\dot{z}_k = \lambda_k z_k + \sum_{i,j} D_{kij} z_i z_j + \boldsymbol{\psi}_k^H \mathbf{B}_c \Delta\mathbf{u}_c + \boldsymbol{\psi}_k^H \mathbf{B}_g \Delta\mathbf{u}_d
\label{eq:rom}
\end{equation}

For $\mathcal{H}_\infty$ control design, the linear ROM (neglecting $D_{kij}$ terms) is used as the design plant. The nonlinear terms are retained in the validation model to assess controller robustness against unmodelled nonlinearities.

\begin{figure}[htbp]
\centering
\includegraphics[width=0.6\textwidth]{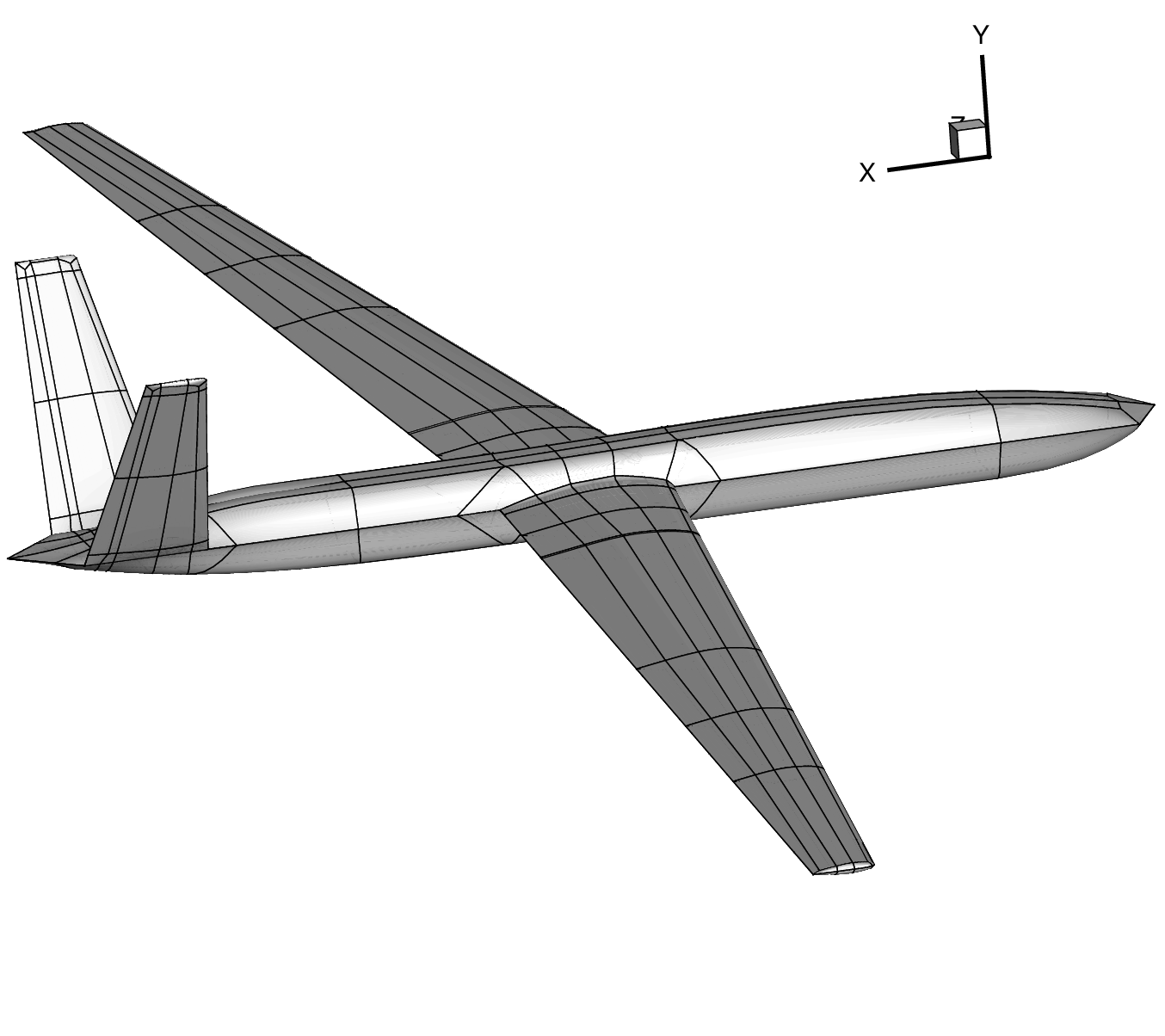}
\caption{Beam finite element model of the Global Hawk-like UAV showing the structural discretisation. The wing (17.75-m semi-span), fuselage, and empennage are modelled with geometrically-exact beam elements. Trailing-edge control surfaces on the outer wing provide the GLA actuation.}
\label{fig:uav_geometry}
\end{figure}

\begin{figure}[htbp]
\centering
\includegraphics[width=0.6\textwidth]{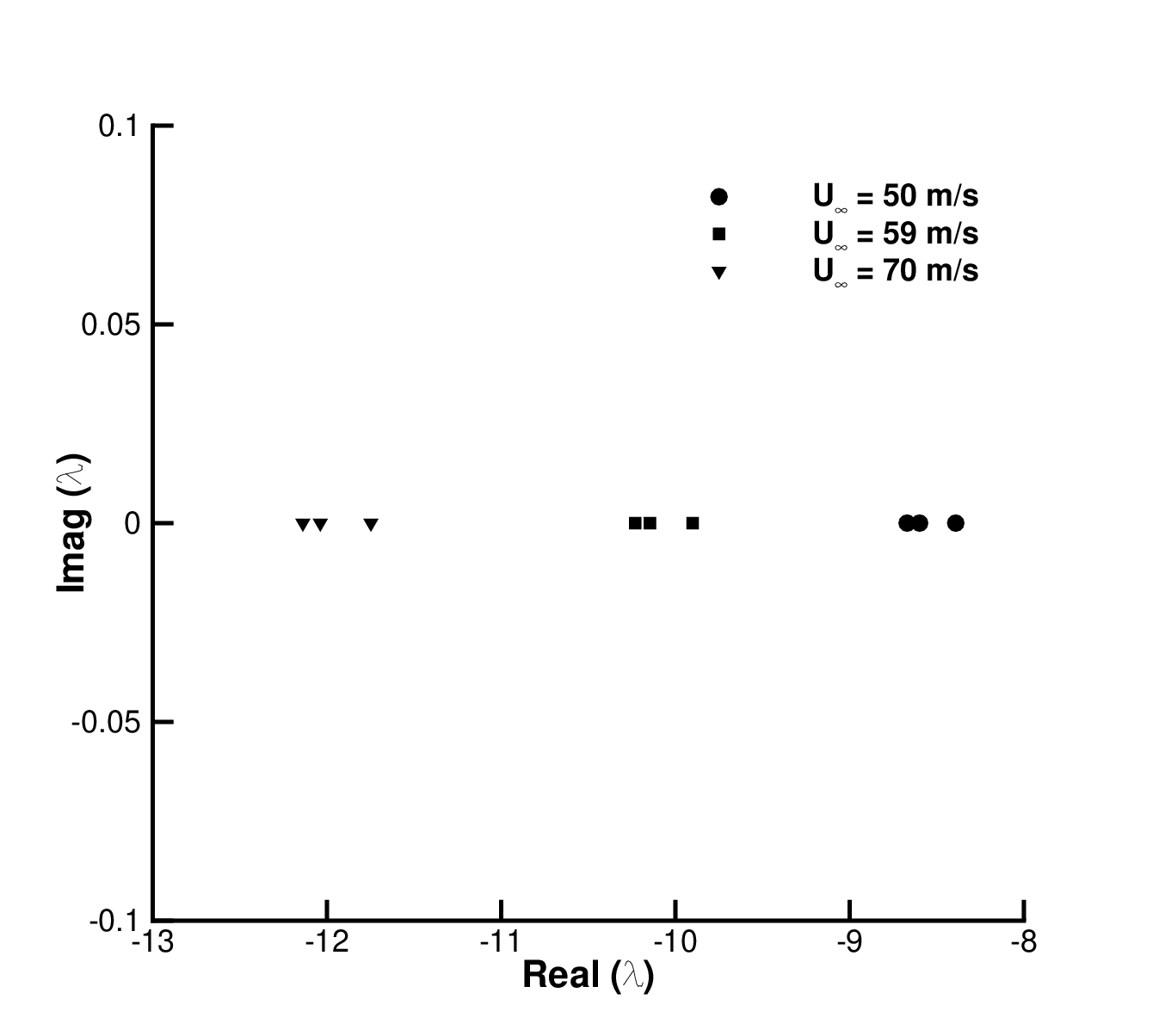}
\caption{Variation of the gust mode eigenvalues of the coupled Jacobian for the UAV with freestream speed ($U_\infty = 50$, 59, 70~m/s). The 3 real gust modes, related to the K\"ussner function constant $\varepsilon_3$, are tracked as the speed changes. The remaining 5 complex conjugate structural modes (bending and torsion) retained in the 8-mode ROM basis lie outside this view due to their larger imaginary parts.}
\label{fig:rom_eigenvalues}
\end{figure}

The UAV test case geometry is shown in~\Cref{fig:uav_geometry}, and the eigenvalue spectrum used for basis selection is presented in~\Cref{fig:rom_eigenvalues}.

\section{$\mathcal{H}_\infty$ Control Design}
\label{sec:hinf}

\subsection{Control Architecture}

\begin{figure}[htbp]
\centering
\includegraphics[width=0.6\textwidth]{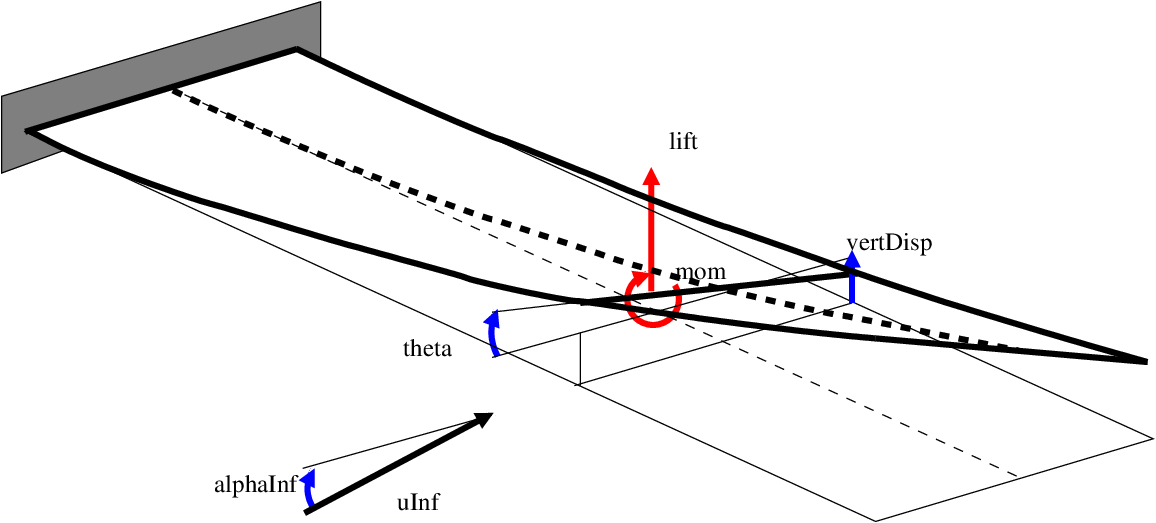}
\caption{Wing deformation under gust loading showing the open-loop (dashed) and closed-loop (solid) responses. The trailing-edge flap, distributed along the outer wing span, provides the GLA actuation by generating a compensating pitching moment.}
\label{fig:wing_deformation}
\end{figure}

The GLA control architecture is shown schematically in~\Cref{fig:wing_deformation}. The measurement output is the wing-root bending moment (or wing-tip displacement), and the control input is the trailing-edge flap angular acceleration $\ddot{\delta}$. The flap dynamics are modelled as a second-order actuator:
\begin{equation}
\dot{\mathbf{x}}_{\text{act}} = \begin{bmatrix} 0 & 1 \\ 0 & 0 \end{bmatrix} \mathbf{x}_{\text{act}} + \begin{bmatrix} 0 \\ 1 \end{bmatrix} u, \qquad \mathbf{x}_{\text{act}} = \begin{bmatrix} \delta \\ \dot{\delta} \end{bmatrix}
\label{eq:actuator}
\end{equation}
where $u = \ddot{\delta}$ is the commanded flap acceleration. This second-order model ensures continuity of both flap position and rate, providing physically realisable commands.

\subsection{Augmented Plant}

The ROM is augmented with the actuator dynamics to form the design plant:
\begin{equation}
\dot{\mathbf{x}} = \mathbf{A}_p \mathbf{x} + \mathbf{B}_w \mathbf{w}_{\text{ext}} + \mathbf{B}_u u
\label{eq:augmented_plant}
\end{equation}
where $\mathbf{x} = \{\mathbf{z}, \delta, \dot{\delta}\}^T$ is the augmented state, $\mathbf{w}_{\text{ext}}$ is the exogenous input (gust), and the plant matrices are:
\begin{equation}
\mathbf{A}_p = \begin{bmatrix} \boldsymbol{\Lambda} & \boldsymbol{\Psi}^H\mathbf{B}_c & \boldsymbol{\Psi}^H\mathbf{B}_{c1} \\ \mathbf{0} & 0 & 1 \\ \mathbf{0} & 0 & 0 \end{bmatrix}, \quad
\mathbf{B}_w = \begin{bmatrix} \boldsymbol{\Psi}^H\mathbf{B}_g \\ 0 \\ 0 \end{bmatrix}, \quad
\mathbf{B}_u = \begin{bmatrix} \boldsymbol{\Psi}^H\mathbf{B}_{c2} \\ 0 \\ 1 \end{bmatrix}
\label{eq:plant_matrices}
\end{equation}
where $\boldsymbol{\Lambda} = \text{diag}(\lambda_1, \ldots, \lambda_m)$ is the diagonal matrix of retained eigenvalues, and $\mathbf{B}_c$, $\mathbf{B}_{c1}$, $\mathbf{B}_{c2}$ are the control derivatives corresponding to flap rotation, angular velocity, and angular acceleration, respectively.

\subsection{$\mathcal{H}_\infty$ Problem Formulation}

The generalised plant is constructed following the standard two-port framework:
\begin{equation}
\begin{bmatrix} \mathbf{z}_{\text{perf}} \\ \mathbf{y}_{\text{meas}} \end{bmatrix} = \begin{bmatrix} \mathbf{P}_{11} & \mathbf{P}_{12} \\ \mathbf{P}_{21} & \mathbf{P}_{22} \end{bmatrix} \begin{bmatrix} \mathbf{w}_{\text{ext}} \\ u \end{bmatrix}
\label{eq:gen_plant}
\end{equation}
where $\mathbf{z}_{\text{perf}}$ is the performance output and $\mathbf{y}_{\text{meas}}$ is the measured output.

An input-shaping weighting function $K_c$ is applied to penalise the control effort:
\begin{equation}
\mathbf{z}_{\text{perf}} = \begin{bmatrix} \mathbf{C}_z \mathbf{x} \\ K_c \cdot u \end{bmatrix}
\label{eq:perf_output}
\end{equation}
where $\mathbf{C}_z$ selects the states to be regulated (typically wing-tip displacement and bending slope).

The $\mathcal{H}_\infty$ optimisation finds a stabilising controller $K(s)$ that minimises the closed-loop $\mathcal{H}_\infty$ norm:
\begin{equation}
\gamma^* = \min_{K \text{ stabilising}} \left\| \mathcal{F}_l(\mathbf{P}, K) \right\|_\infty = \min_{K} \sup_\omega \bar{\sigma}\left[ \mathcal{F}_l(\mathbf{P}, K)(j\omega) \right]
\label{eq:hinf_opt}
\end{equation}
where $\mathcal{F}_l$ denotes the lower linear fractional transformation and $\bar{\sigma}$ is the maximum singular value. The optimal $\gamma^*$ represents the minimum achievable worst-case amplification from gust to performance output.

\subsection{Role of the Weighting Parameter $K_c$}
\label{sec:kc_role}

The weighting parameter $K_c$ penalises control effort: larger values of $K_c$ produce smaller flap deflections at the expense of reduced gust alleviation. The value $K_c = 1$ was used for all results presented in this paper, following~\citet{Tantaroudas2017bookchapter}.

\section{Results}
\label{sec:results}

\subsection{Global Hawk-Like UAV}
\label{sec:uav_results}

\begin{figure}[htbp]
\centering
\includegraphics[width=0.6\textwidth]{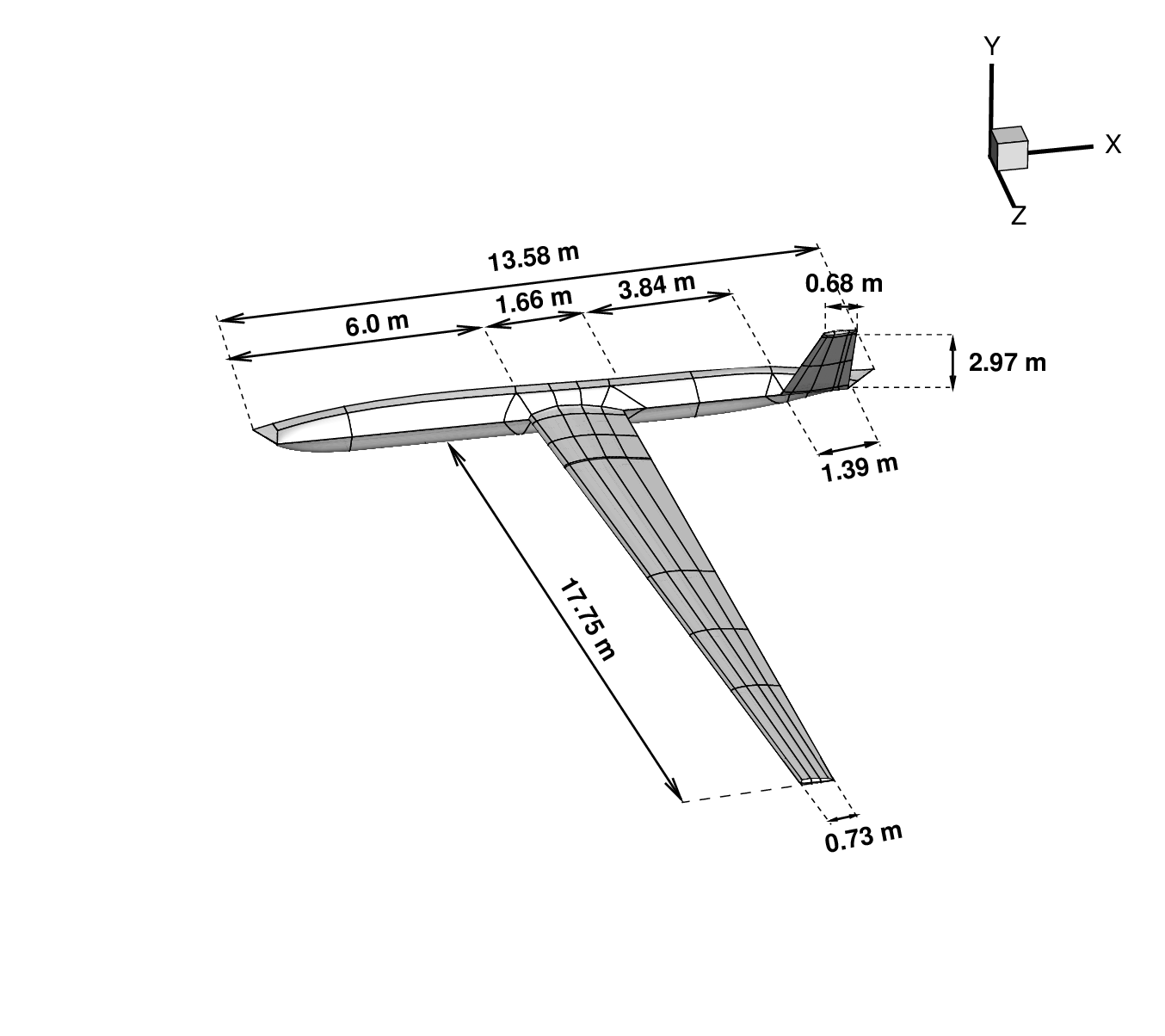}
\caption{Three-view geometry of the Global Hawk-like UAV showing the wing, fuselage, horizontal stabiliser, and vertical tail. The trailing-edge control surfaces used for GLA are located on the outer wing sections.}
\label{fig:dstl_geometry}
\end{figure}

\begin{figure}[htbp]
\centering
\includegraphics[width=0.55\textwidth]{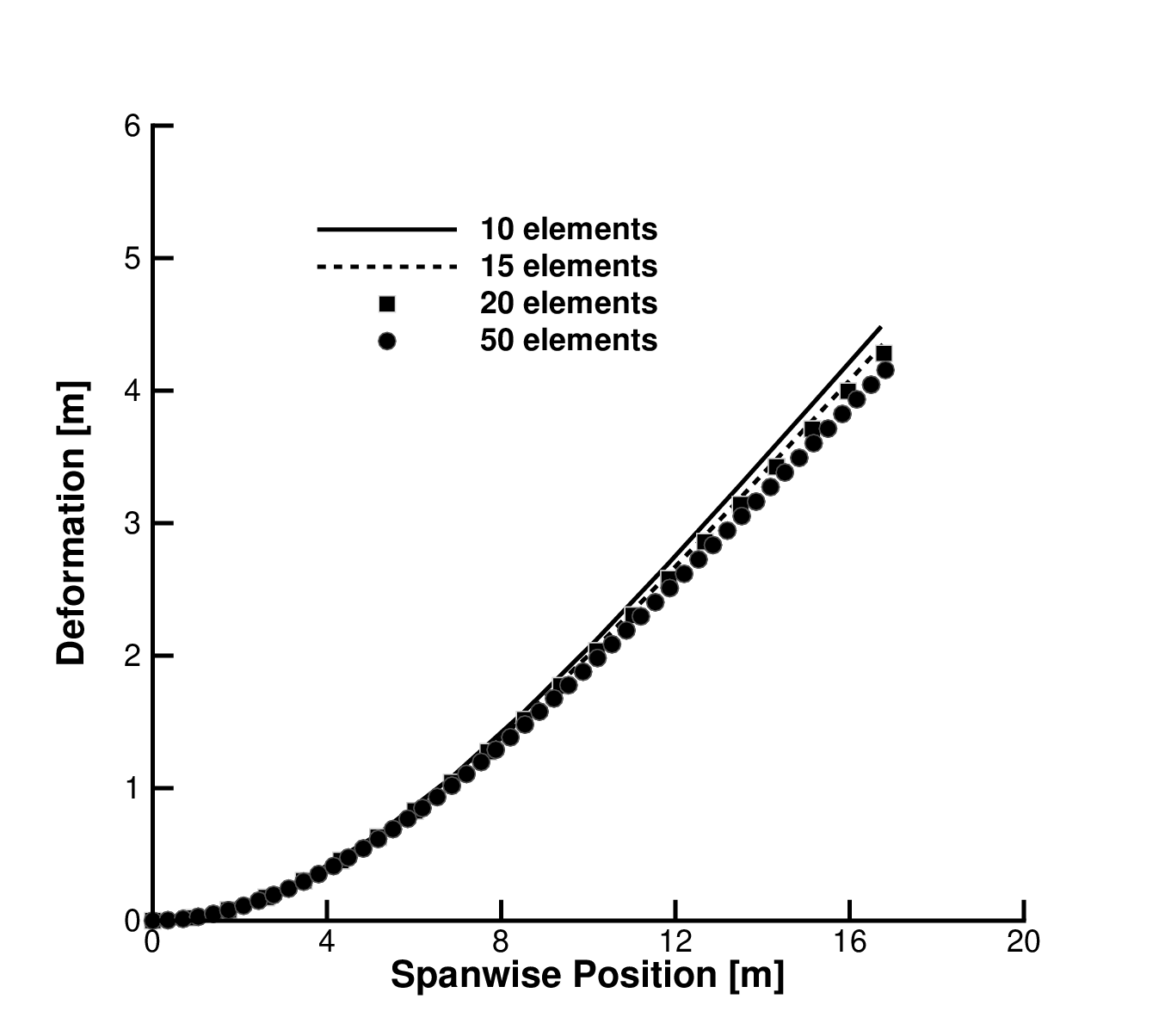}
\caption{Mesh convergence of the static aeroelastic wing deformation of the UAV at the trim flight condition ($U_\infty = 59$~m/s, $\alpha_0 = 4^\circ$) for 10, 15, 20, and 50 beam elements. Convergence is achieved with 20 elements, which is used as the standard discretisation. The converged equilibrium deformation serves as the reference state for both the ROM construction and the $\mathcal{H}_\infty$ controller design.}
\label{fig:static_deformation}
\end{figure}

The UAV has a 17.75-m semi-span wing modelled with 20 beam elements, yielding $n = 540$ full-order DOF reduced to $m = 8$ ROM modes~\citep{Tantaroudas2014aviation, Tantaroudas2017bookchapter}. The DSTL (Defence Science and Technology Laboratory) geometry representation is shown in~\Cref{fig:dstl_geometry}. Flight conditions: $U_\infty = 59$~m/s, $\rho = 0.0789$~kg/m$^3$, trim angle of attack $\alpha_0 = 4^\circ$. The static aeroelastic deformation at the trim state is presented in~\Cref{fig:static_deformation}.

\begin{figure}[htbp]
\centering
\begin{subfigure}[b]{0.48\textwidth}
\includegraphics[width=\textwidth]{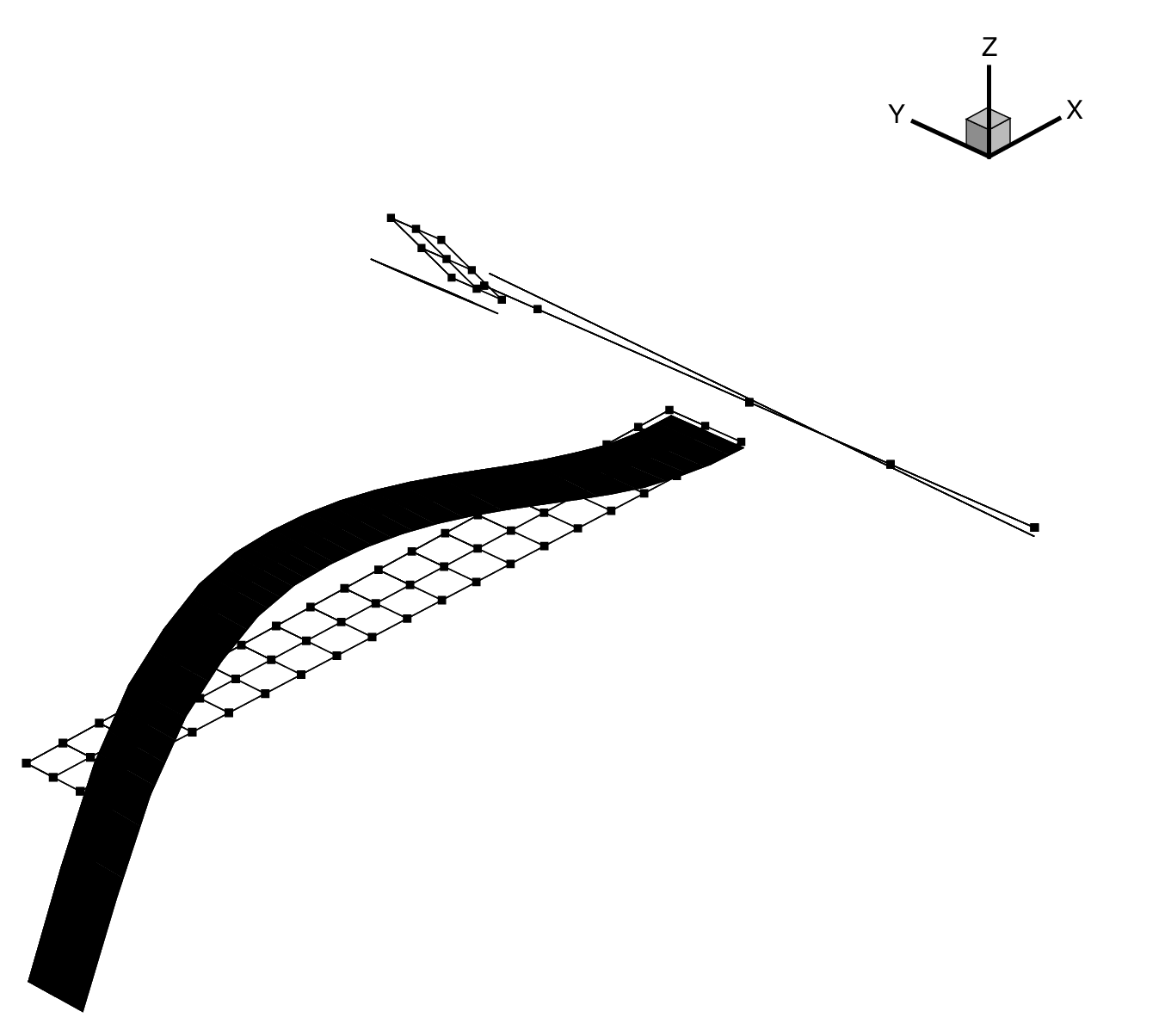}
\caption{First wing bending mode}
\label{fig:beam_mode1}
\end{subfigure}
\hfill
\begin{subfigure}[b]{0.48\textwidth}
\includegraphics[width=\textwidth]{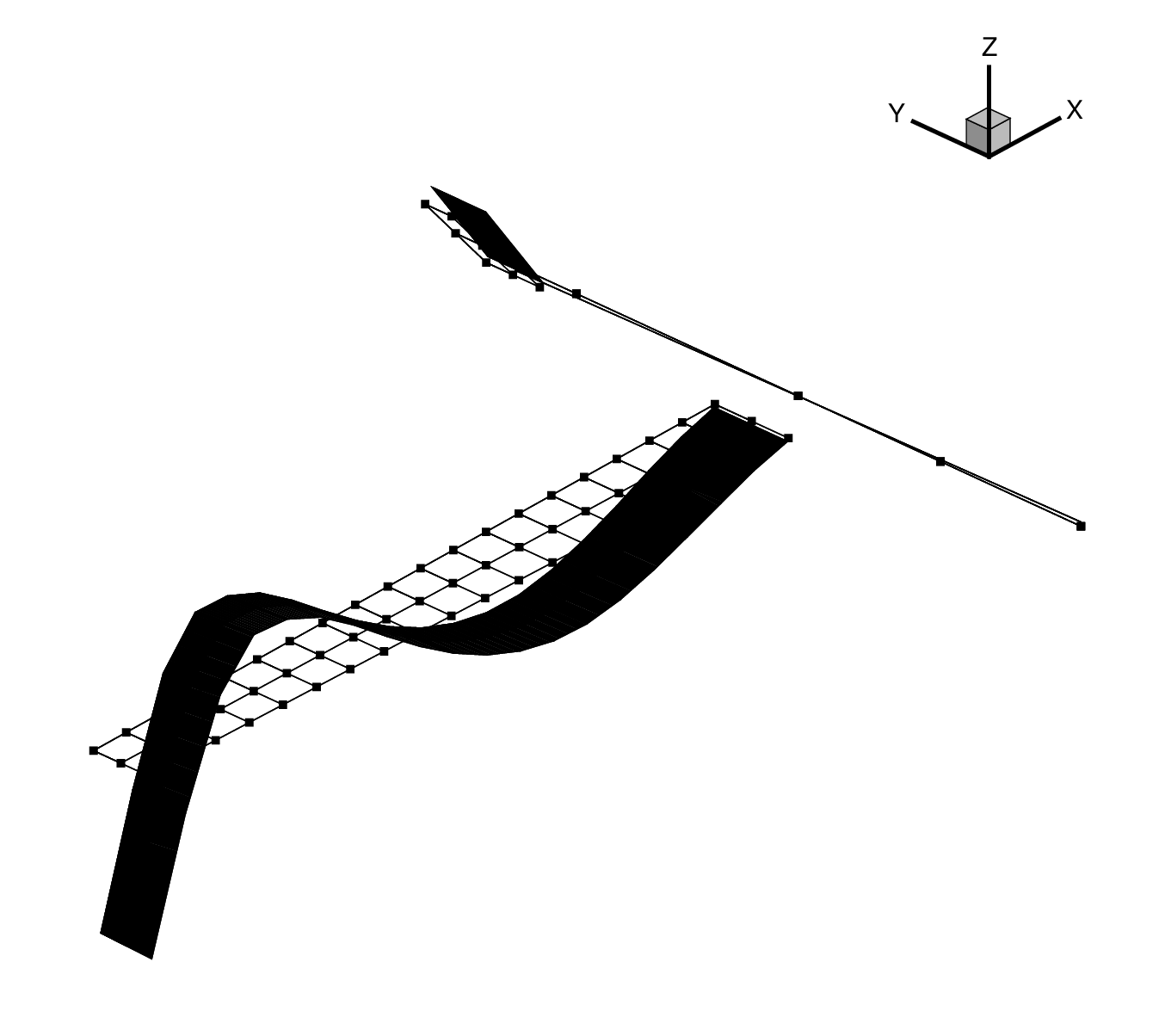}
\caption{Second wing bending mode}
\label{fig:beam_mode2}
\end{subfigure}
\caption{Structural mode shapes of the UAV wing obtained from the linearised beam model. These modes contribute most significantly to the gust response and are the primary targets of the $\mathcal{H}_\infty$ GLA controller.}
\label{fig:beam_modes}
\end{figure}

The first two wing bending mode shapes, which dominate the gust response, are shown in~\Cref{fig:beam_modes}.

\subsubsection{Discrete gust response}

\begin{figure}[htbp]
\centering
\begin{subfigure}[b]{0.48\textwidth}
\includegraphics[width=\textwidth]{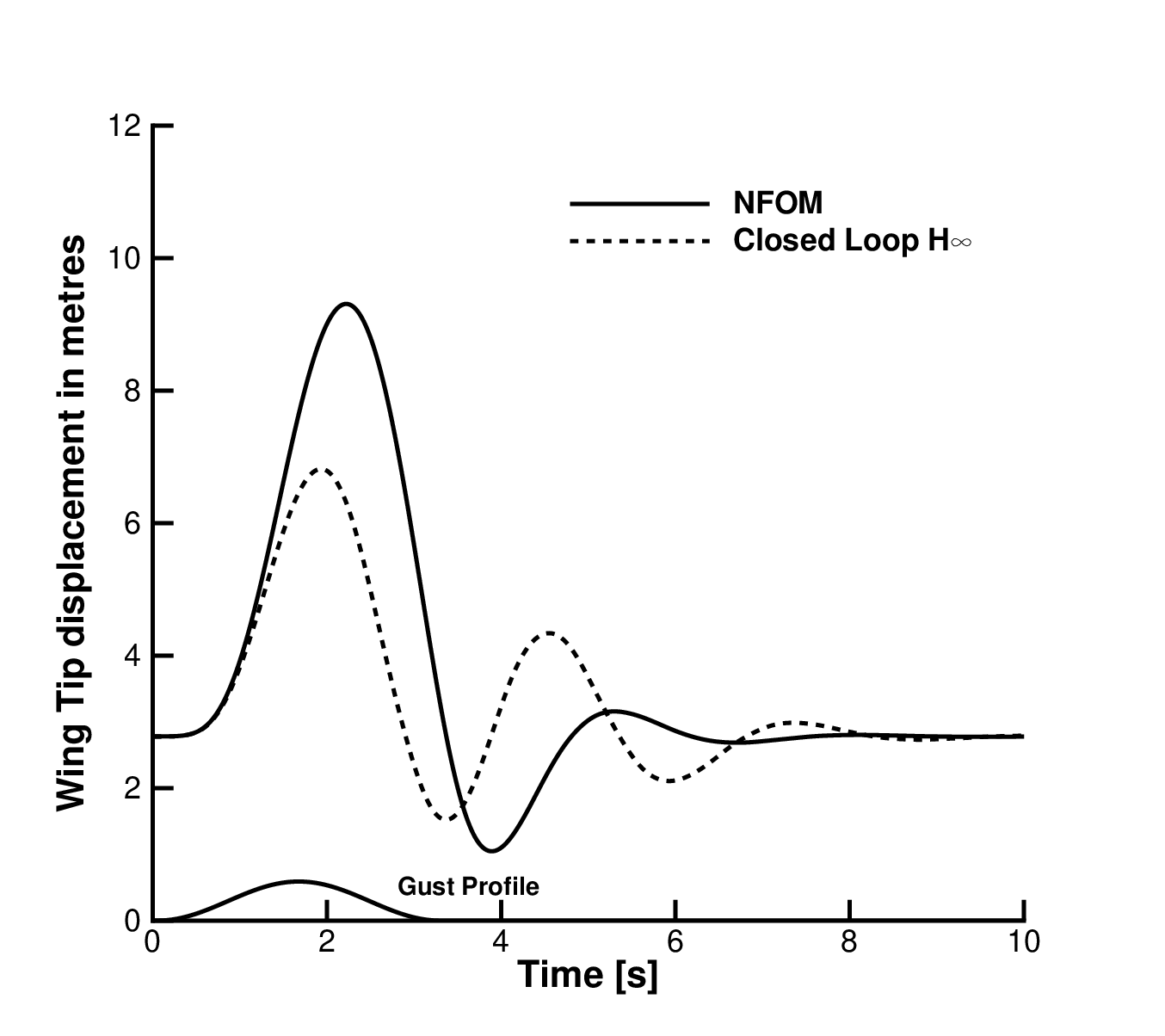}
\caption{Wing-tip vertical displacement}
\label{fig:wtip_hinf_cos}
\end{subfigure}
\hfill
\begin{subfigure}[b]{0.48\textwidth}
\includegraphics[width=\textwidth]{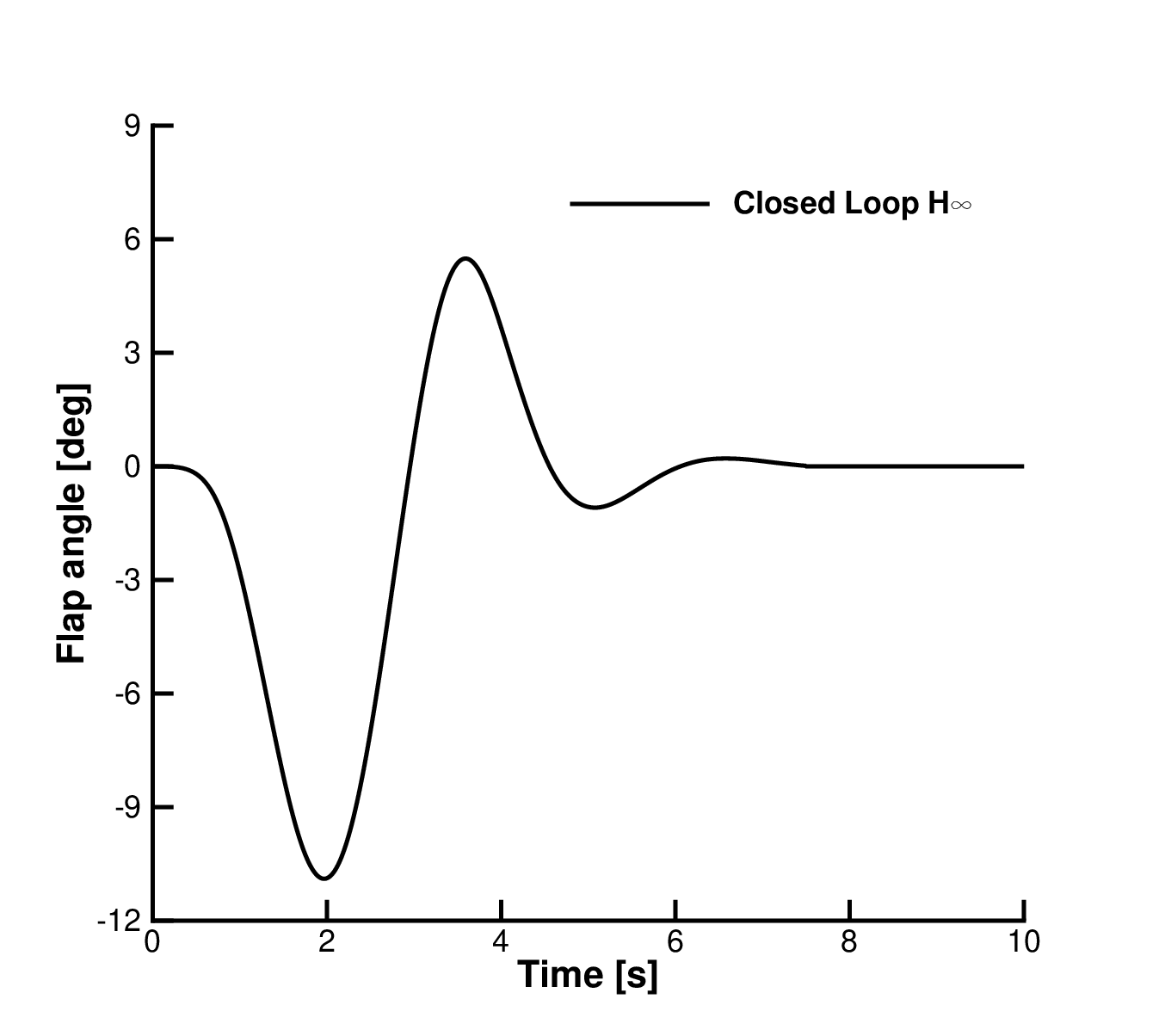}
\caption{Trailing-edge flap deflection}
\label{fig:flap_hinf_cos}
\end{subfigure}
\caption{$\mathcal{H}_\infty$ GLA performance under worst-case discrete gust (UAV, $K_c = 1$): open loop (solid black) and closed loop (dashed red). The controller achieves 23.15\% reduction in peak wing-tip deflection with a maximum flap rotation of $-9.47^\circ$.}
\label{fig:hinf_cos_results}
\end{figure}

The worst-case ``1-minus-cosine'' discrete gust has a duration of approximately 4~s (197~MAC) and peak velocity corresponding to 14\% of the freestream speed. \Cref{fig:hinf_cos_results} presents the closed-loop wing-tip displacement and flap deflection time histories.

\begin{table}[htbp]
\centering
\caption{$\mathcal{H}_\infty$ GLA performance metrics under discrete gust (UAV, $K_c = 1$).}
\label{tab:hinf_discrete}
\begin{tabular}{@{}lcc@{}}
\toprule
Metric & Open loop & Closed loop ($K_c = 1$) \\
\midrule
Wing-tip deflection reduction & --- & $-23.15\%$ \\
Max flap rotation & --- & $-9.47^\circ$ \\
\bottomrule
\end{tabular}
\end{table}

The controller commands an initial downward flap rotation, which generates a nose-down pitching moment that partially counteracts the gust-induced lift increment. The flap deflection profile exhibits a smooth bell-shaped response, consistent with the second-order actuator model in~\Cref{eq:actuator}. The closed-loop wing-tip response shows both reduced peak amplitude and enhanced damping of the post-gust oscillation.

\subsubsection{Stochastic turbulence response}

\begin{figure}[htbp]
\centering
\begin{subfigure}[b]{0.48\textwidth}
\includegraphics[width=\textwidth]{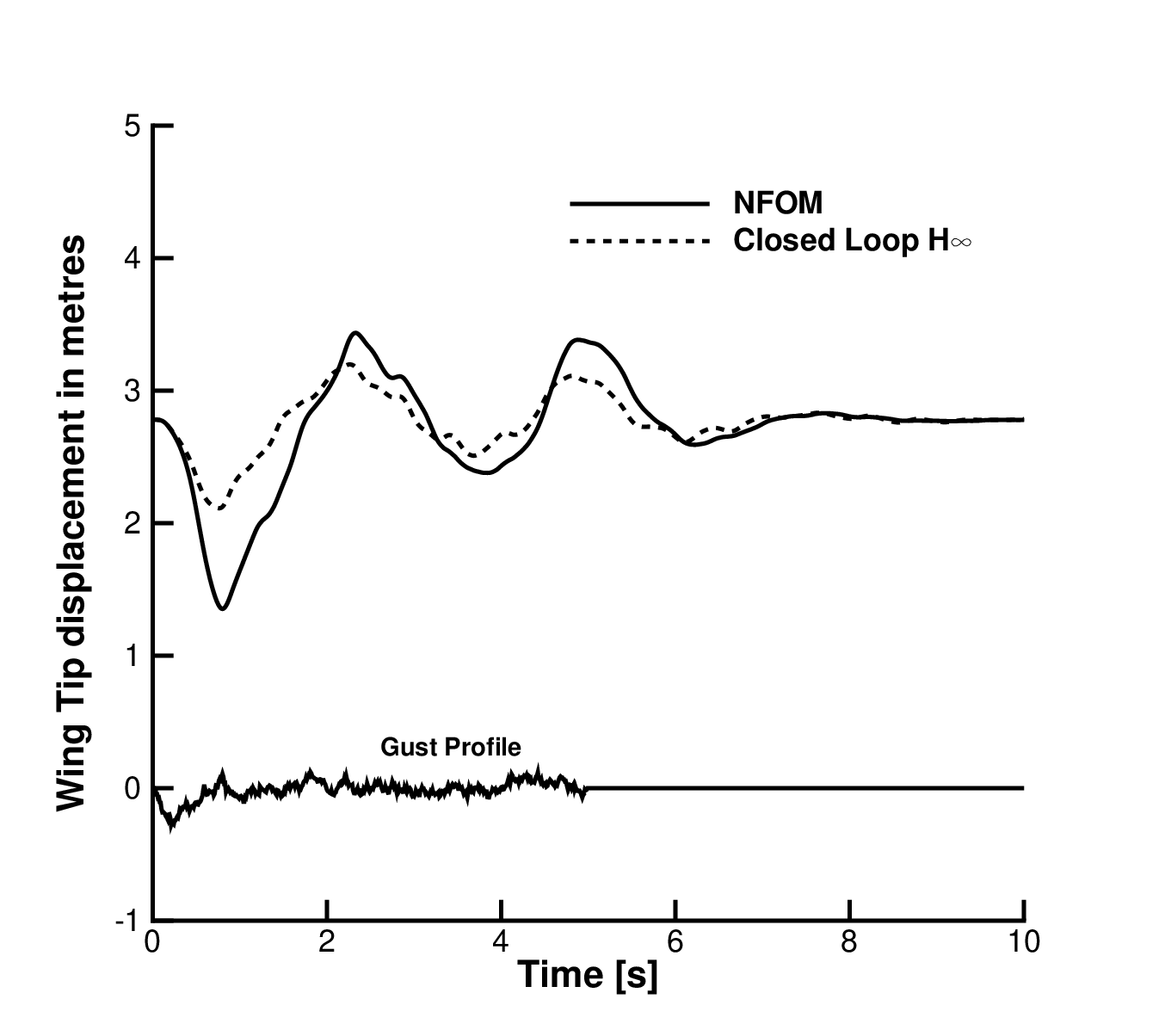}
\caption{Wing-tip vertical displacement}
\label{fig:wtip_hinf_vk}
\end{subfigure}
\hfill
\begin{subfigure}[b]{0.48\textwidth}
\includegraphics[width=\textwidth]{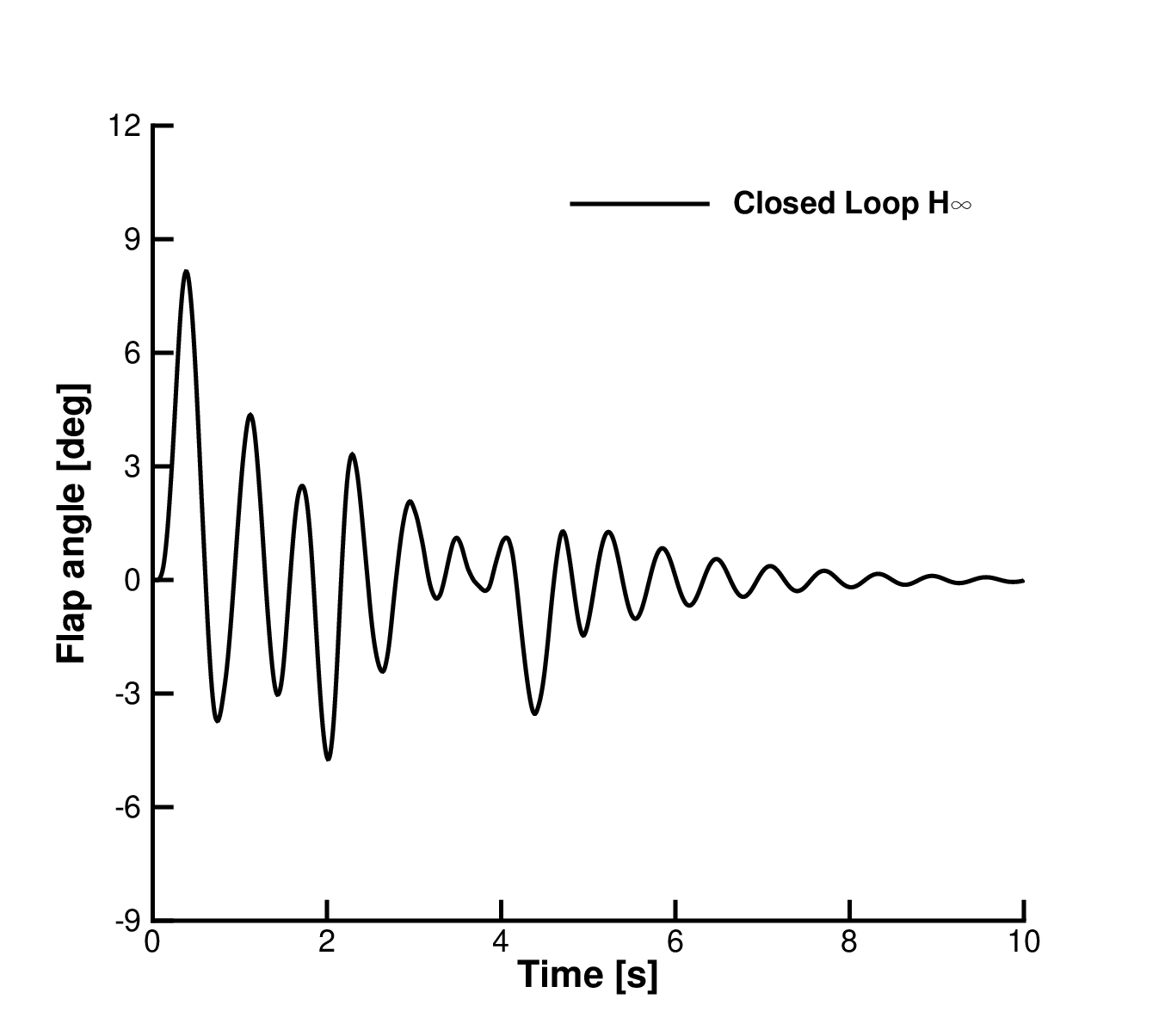}
\caption{Trailing-edge flap deflection}
\label{fig:flap_hinf_vk}
\end{subfigure}
\caption{$\mathcal{H}_\infty$ GLA performance under Von K\'arm\'an stochastic turbulence (UAV, $K_c = 1$): open loop (solid black) and closed loop (dashed red). The RMS wing-tip displacement is reduced by 10.26\%.}
\label{fig:hinf_vk_results}
\end{figure}

Under Von K\'arm\'an continuous turbulence ($L_w = 750$~m), the closed-loop performance is presented in~\Cref{fig:hinf_vk_results}.

\begin{table}[htbp]
\centering
\caption{$\mathcal{H}_\infty$ GLA performance under Von K\'arm\'an turbulence (UAV, $K_c = 1$).}
\label{tab:hinf_vk}
\begin{tabular}{@{}lcc@{}}
\toprule
Metric & Open loop & Closed loop ($K_c = 1$) \\
\midrule
RMS wing-tip deflection & baseline & $-10.26\%$ \\
Max flap rotation & --- & $\pm 12.79^\circ$ \\
\bottomrule
\end{tabular}
\end{table}

The reduced performance under stochastic turbulence (10.26\% RMS reduction) compared to discrete gust (23.15\% peak reduction) reflects the broadband frequency content of the Von K\'arm\'an spectrum: the controller effectively attenuates energy within its bandwidth but cannot reject high-frequency content that excites modes beyond the 8-mode ROM~\citep{DaRonch2013gust}.

\subsection{Model Convergence and Extension}

\begin{figure}[htbp]
\centering
\begin{subfigure}[b]{0.48\textwidth}
\includegraphics[width=\textwidth]{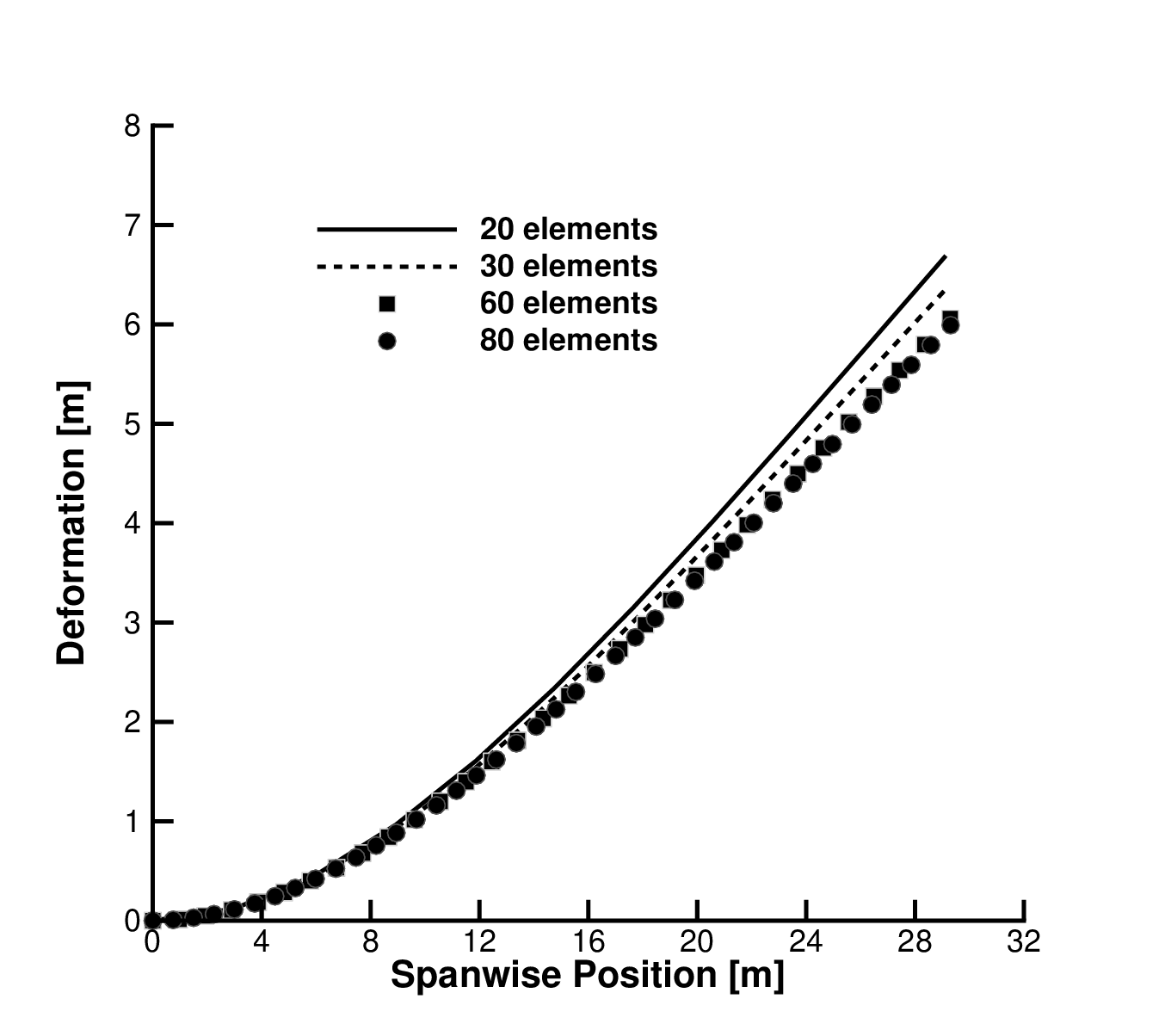}
\caption{Static deformation convergence (VFA flying-wing)}
\label{fig:static_convergence}
\end{subfigure}
\hfill
\begin{subfigure}[b]{0.48\textwidth}
\includegraphics[width=\textwidth]{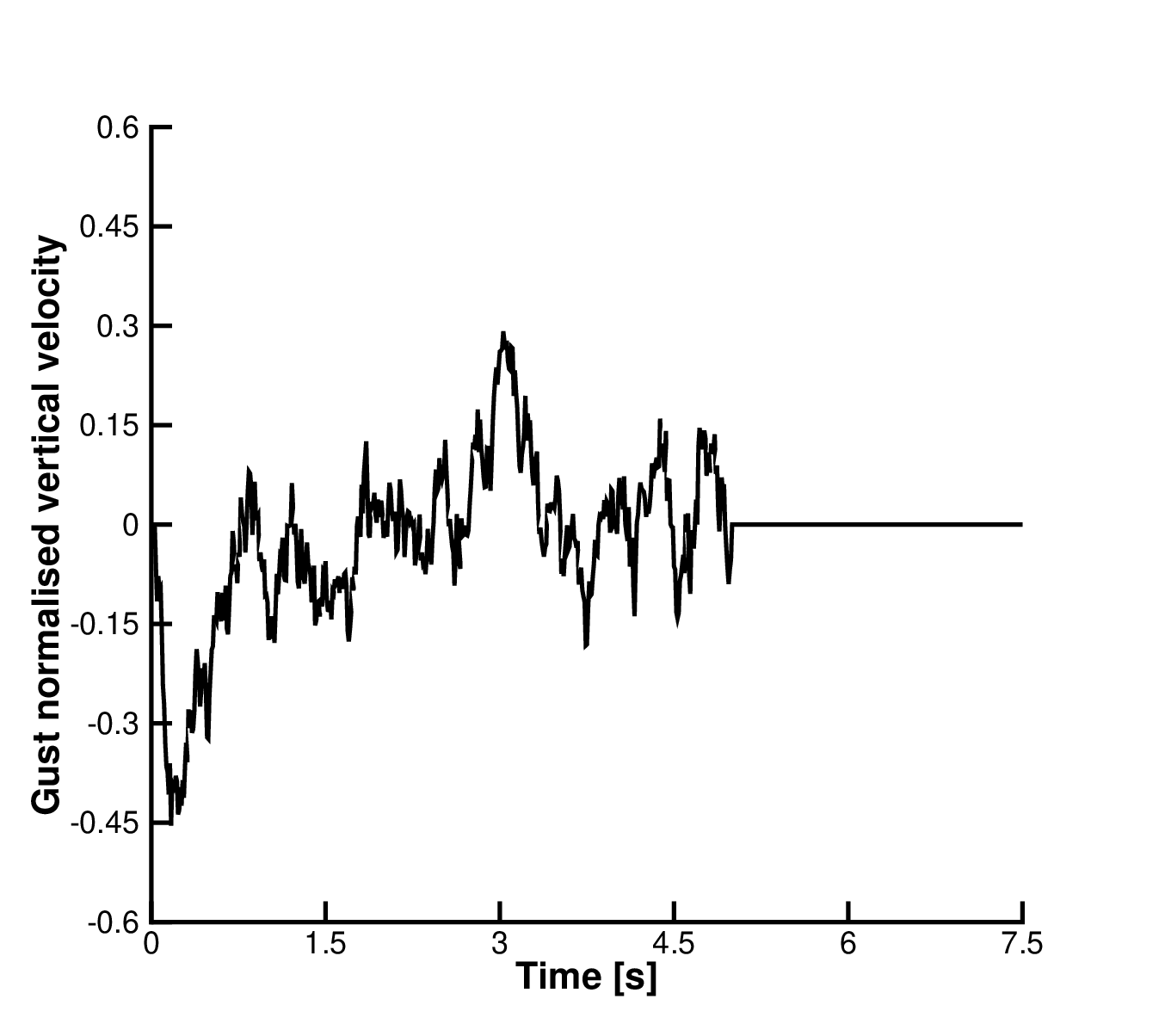}
\caption{Von K\'arm\'an turbulence profile}
\label{fig:vk_gust_convergence}
\end{subfigure}
\caption{(a)~Mesh convergence of the nonlinear static wing deformation for the very flexible flying-wing (32-m span) with increasing number of beam elements (20, 30, 60, 80), and (b)~sample Von K\'arm\'an turbulence velocity profile (normalised by freestream speed) used as the stochastic gust input.}
\label{fig:convergence}
\end{figure}

\Cref{fig:static_convergence} demonstrates the mesh convergence of the static aeroelastic solution for the very flexible flying-wing (VFA) configuration (32-m span, 16-m semi-span, $n = 1{,}616$ DOF, $m = 9$ ROM modes), confirming that 80 beam elements adequately resolve the nonlinear wing deformation for this larger configuration. \Cref{fig:vk_gust_convergence} shows a representative Von K\'arm\'an turbulence velocity profile (normalised by the freestream speed) employed in the stochastic gust simulations presented in~\Cref{sec:uav_results}.

\section{Discussion}
\label{sec:discussion}

\textbf{Robustness of ROM-based design.} The $\mathcal{H}_\infty$ controller, designed on the 8-mode linear ROM, provides consistent load alleviation when applied to the 540-DOF nonlinear full-order model. This robustness arises from the inherent conservatism of $\mathcal{H}_\infty$ synthesis, which accounts for worst-case disturbance amplification and therefore provides implicit robustness against the unmodelled nonlinearities present in the full-order system.

\textbf{Discrete versus stochastic performance.} The controller achieves 23.15\% reduction under discrete gusts versus 10.26\% under stochastic turbulence. This asymmetry has a clear physical explanation: discrete gusts concentrate energy at a single frequency ($f_g = U_\infty/H_g$) that falls within the controller bandwidth, enabling effective attenuation. Stochastic turbulence distributes energy across a broad spectrum, and the controller can only attenuate the portion within its bandwidth. For improved stochastic performance, adaptive controllers that can adjust their response in real time may be more effective~\citep{Tantaroudas2014aviation}.

\textbf{Practical actuator constraints.} The maximum flap rotation of $\sim$10$^\circ$ at $K_c = 1$ is within the typical authority of trailing-edge control surfaces on UAVs. However, the current model does not include explicit actuator rate and magnitude limits, nor does it account for hinge moment constraints. In a detailed design, these constraints would be incorporated as additional weighting functions or as hard constraints in a mixed $\mathcal{H}_2/\mathcal{H}_\infty$ formulation.

\textbf{Extension to the VFA flying-wing.} The same methodology was applied to the 32-m span very flexible flying-wing ($n = 1{,}616$ DOF, $m = 9$ ROM modes) in~\citep{Tantaroudas2015scitech}, where the controller successfully reduced wing-tip deflections while maintaining the trim flight condition. The mesh convergence for this configuration is shown in~\Cref{fig:static_convergence}. The larger deformations of this configuration introduce stronger nonlinear effects, but the $\mathcal{H}_\infty$ controller's inherent robustness margin accommodates these without degradation.

\section{Conclusions}
\label{sec:conclusions}

An $\mathcal{H}_\infty$ robust control framework for gust load alleviation of very flexible aircraft has been presented. The controller is designed on a compact reduced-order model and validated on the full nonlinear system. The $\mathcal{H}_\infty$ controller designed on an 8-mode ROM successfully alleviates gust loads on the 540-DOF nonlinear full-order model, achieving 23.15\% wing-tip deflection reduction under worst-case discrete gust and 10.26\% under stochastic Von K\'arm\'an turbulence. The input-shaping weighting parameter $K_c$ provides a systematic single-parameter trade-off between structural load alleviation and flight path integrity, with $K_c = 1$ used for the configurations studied following~\citet{Tantaroudas2017bookchapter}. Better performance is obtained for longer-duration discrete gusts whose frequency content falls within the controller bandwidth, consistent with the frequency-domain interpretation of $\mathcal{H}_\infty$ optimality. The framework is directly extensible to larger, more flexible configurations (1,616-DOF flying-wing) and to higher-fidelity aerodynamic models~\citep{DaRonch2012rom, Tantaroudas2017bookchapter}. For improved stochastic turbulence attenuation, adaptive control strategies may complement the $\mathcal{H}_\infty$ design~\citep{Tantaroudas2014aviation}.

\section*{Acknowledgements}

This work was supported by the U.K.\ Engineering and Physical Sciences Research Council (EPSRC) grant EP/I014594/1 on ``Nonlinear Flexibility Effects on Flight Dynamics and Control of Next-Generation Aircraft.'' The authors are grateful to Prof.\ A.\ Da Ronch and Prof.\ K.J.\ Badcock for their valuable guidance on flight dynamics modelling.

\bibliographystyle{unsrtnat}
\bibliography{references}

@INBOOK{Tantaroudas2017bookchapter,
  author    = {Tantaroudas, N.D. and Da Ronch, A.},
  title     = {Nonlinear Reduced-order Aeroservoelastic Analysis of Very Flexible Aircraft},
  booktitle = {Advanced UAV Aerodynamics, Flight Stability and Control},
  publisher = {John Wiley \& Sons, Ltd},
  address   = {Chichester, UK},
  chapter   = {4},
  pages     = {143--179},
  year      = {2017},
  doi       = {10.1002/9781118928691.ch4}
}

@INPROCEEDINGS{Tantaroudas2015scitech,
  author    = {Tantaroudas, N.D. and Da Ronch, A. and Badcock, K.J. and Palacios, R.},
  title     = {Model Order Reduction for Control Design of Flexible Free-Flying Aircraft},
  booktitle = {AIAA Atmospheric Flight Mechanics Conference, AIAA SciTech 2015},
  series    = {AIAA Paper 2015-0240},
  year      = {2015},
  doi       = {10.2514/6.2015-0240}
}

@INPROCEEDINGS{Tantaroudas2014aviation,
  author    = {Tantaroudas, N.D. and Da Ronch, A. and Gai, G. and Badcock, K.J. and Palacios, R.},
  title     = {An Adaptive Aeroelastic Control Approach using Non Linear Reduced Order Models},
  booktitle = {14th AIAA Aviation Technology, Integration, and Operations Conference},
  series    = {AIAA Paper 2014-2590},
  year      = {2014},
  doi       = {10.2514/6.2014-2590}
}

@INPROCEEDINGS{DaRonch2014scitech_flight,
  author    = {Da Ronch, A. and McCracken, A.J. and Tantaroudas, N.D. and Badcock, K.J. and Hesse, H. and Palacios, R.},
  title     = {Assessing the Impact of Aerodynamic Modelling on Manoeuvring Aircraft},
  booktitle = {AIAA SciTech 2014, AIAA Atmospheric Flight Mechanics Conference},
  series    = {AIAA Paper 2014-0732},
  year      = {2014},
  doi       = {10.2514/6.2014-0732}
}

@INPROCEEDINGS{DaRonch2014flutter,
  author    = {Da Ronch, A. and Tantaroudas, N.D. and Jiffri, S. and Mottershead, J.E.},
  title     = {A Nonlinear Controller for Flutter Suppression: from Simulation to Wind Tunnel Testing},
  booktitle = {55th AIAA/ASME/ASCE/AHS/SC Structures, Structural Dynamics, and Materials Conference},
  series    = {AIAA Paper 2014-0345},
  year      = {2014},
  doi       = {10.2514/6.2014-0345}
}

@INPROCEEDINGS{Fichera2014isma,
  author    = {Fichera, S. and Jiffri, S. and Wei, X. and Da Ronch, A. and Tantaroudas, N.D. and Mottershead, J.E.},
  title     = {Experimental and Numerical Study of Nonlinear Dynamic Behaviour of an Aerofoil},
  booktitle = {ISMA 2014 Conference on Noise and Vibration Engineering},
  pages     = {3609--3618},
  year      = {2014}
}

@INPROCEEDINGS{DaRonch2013gust,
  author    = {Da Ronch, A. and Tantaroudas, N.D. and Timme, S. and Badcock, K.J.},
  title     = {Model Reduction for Linear and Nonlinear Gust Loads Analysis},
  booktitle = {54th AIAA/ASME/ASCE/AHS/ASC Structures, Structural Dynamics, and Materials Conference},
  series    = {AIAA Paper 2013-1492},
  year      = {2013},
  doi       = {10.2514/6.2013-1492}
}

@INPROCEEDINGS{DaRonch2013control,
  author    = {Da Ronch, A. and Tantaroudas, N.D. and Badcock, K.J.},
  title     = {Reduction of Nonlinear Models for Control Applications},
  booktitle = {54th AIAA/ASME/ASCE/AHS/ASC Structures, Structural Dynamics, and Materials Conference},
  series    = {AIAA Paper 2013-1491},
  year      = {2013},
  doi       = {10.2514/6.2013-1491}
}

@INPROCEEDINGS{Papatheou2013ifasd,
  author    = {Papatheou, E. and Tantaroudas, N.D. and Da Ronch, A. and Cooper, J.E. and Mottershead, J.E.},
  title     = {Active Control for Flutter Suppression: an Experimental Investigation},
  booktitle = {International Forum on Aeroelasticity and Structural Dynamics},
  series    = {IFASD Paper 2013-8D},
  year      = {2013}
}

@ARTICLE{Patil2001,
  author  = {Patil, M.J. and Hodges, D.H. and Cesnik, C.E.S.},
  title   = {Nonlinear Aeroelasticity and Flight Dynamics of High-Altitude Long-Endurance Aircraft},
  journal = {Journal of Aircraft},
  volume  = {38},
  number  = {1},
  pages   = {88--94},
  year    = {2001},
  doi     = {10.2514/2.2738}
}

@ARTICLE{Patil2006,
  author  = {Patil, M.J. and Hodges, D.H.},
  title   = {Flight Dynamics of Highly Flexible Flying Wings},
  journal = {Journal of Aircraft},
  volume  = {43},
  number  = {6},
  pages   = {1790--1799},
  year    = {2006},
  doi     = {10.2514/1.17640}
}

@INPROCEEDINGS{DaRonch2012rom,
  author    = {Da Ronch, A. and Badcock, K.J. and Wang, Y. and Wynn, A. and Palacios, R.},
  title     = {Nonlinear Model Reduction for Flexible Aircraft Control Design},
  booktitle = {AIAA Atmospheric Flight Mechanics Conference},
  series    = {AIAA Paper 2012-4404},
  year      = {2012},
  doi       = {10.2514/6.2012-4404}
}

@ARTICLE{Theodorsen1935,
  author  = {Theodorsen, T.},
  title   = {General Theory of Aerodynamic Instability and the Mechanism of Flutter},
  journal = {NACA Report},
  volume  = {496},
  year    = {1935}
}

@ARTICLE{Wagner1925,
  author  = {Wagner, H.},
  title   = {{\"U}ber die Entstehung des dynamischen Auftriebes von Tragfl{\"u}geln},
  journal = {Zeitschrift f{\"u}r Angewandte Mathematik und Mechanik},
  volume  = {5},
  number  = {1},
  pages   = {17--35},
  year    = {1925}
}

@ARTICLE{Kussner1936,
  author  = {K{\"u}ssner, H.G.},
  title   = {Zusammenfassender Bericht {\"u}ber den instation{\"a}ren Auftrieb von Fl{\"u}geln},
  journal = {Luftfahrtforschung},
  volume  = {13},
  pages   = {410--424},
  year    = {1936}
}

@ARTICLE{Palacios2010,
  author  = {Palacios, R.},
  title   = {Nonlinear Normal Modes in an Intrinsic Theory of Anisotropic Beams},
  journal = {Journal of Sound and Vibration},
  volume  = {330},
  number  = {8},
  pages   = {1772--1792},
  year    = {2011},
  doi     = {10.1016/j.jsv.2010.10.023}
}

@ARTICLE{Hodges2003,
  author  = {Hodges, D.H.},
  title   = {Geometrically Exact, Intrinsic Theory for Dynamics of Curved and Twisted Anisotropic Beams},
  journal = {AIAA Journal},
  volume  = {41},
  number  = {6},
  pages   = {1131--1137},
  year    = {2003},
  doi     = {10.2514/2.2054}
}

@ARTICLE{Livne2018,
  author  = {Livne, E.},
  title   = {Aircraft Active Flutter Suppression: State of the Art and Technology Maturation Needs},
  journal = {Journal of Aircraft},
  volume  = {55},
  number  = {1},
  pages   = {410--452},
  year    = {2018},
  doi     = {10.2514/1.C034442}
}

@ARTICLE{Noll2004,
  author  = {Noll, T.E. and Brown, J.M. and Perez-Davis, M.E. and Ishmael, S.D. and Tiffany, G.C. and Gaier, M.},
  title   = {Investigation of the {Helios} Prototype Aircraft Mishap},
  journal = {NASA Report},
  year    = {2004}
}

@ARTICLE{Wang2016aiaa,
  author  = {Wang, Y. and Wynn, A. and Palacios, R.},
  title   = {Nonlinear Modal Aeroservoelastic Analysis Framework for Flexible Aircraft},
  journal = {AIAA Journal},
  volume  = {54},
  number  = {10},
  pages   = {3075--3090},
  year    = {2016},
  doi     = {10.2514/1.J054537}
}

@ARTICLE{Wang2018joa,
  author  = {Wang, Y. and Wynn, A. and Palacios, R.},
  title   = {Nonlinear Aeroelastic Control of Very Flexible Aircraft Using Model Updating},
  journal = {Journal of Aircraft},
  volume  = {55},
  number  = {4},
  pages   = {1551--1563},
  year    = {2018},
  doi     = {10.2514/1.C034684}
}

@ARTICLE{Waitman2020,
  author  = {Waitman, S. and Marcos, A.},
  title   = {$\mathcal{H}_\infty$ Control Design for Active Flutter Suppression of Flexible-Wing Unmanned Aerial Vehicle Demonstrator},
  journal = {Journal of Guidance, Control, and Dynamics},
  volume  = {43},
  number  = {4},
  pages   = {656--672},
  year    = {2020},
  doi     = {10.2514/1.G004618}
}

@ARTICLE{Artola2021,
  author  = {Artola, M. and Goizueta, N. and Wynn, A. and Palacios, R.},
  title   = {Aeroelastic Control and Estimation with a Minimal Nonlinear Modal Description},
  journal = {AIAA Journal},
  volume  = {59},
  number  = {7},
  pages   = {2697--2713},
  year    = {2021},
  doi     = {10.2514/1.J060018}
}

@ARTICLE{Tantaroudas2026nmor,
  author  = {Tantaroudas, N.D. and Da Ronch, A. and Karachalios, I. and Badcock, K.J.},
  title   = {Nonlinear Model Order Reduction for Coupled Aeroelastic-Flight Dynamic Systems},
  journal = {arXiv preprint arXiv:2603.15296},
  year    = {2026}
}

@ARTICLE{Tantaroudas2026gust,
  author  = {Tantaroudas, N.D. and Da Ronch, A. and Karachalios, I. and Badcock, K.J.},
  title   = {Rapid Worst-Case Gust Identification for Very Flexible Aircraft Using Reduced-Order Models},
  journal = {arXiv preprint arXiv:2603.16212},
  year    = {2026}
}

@BOOK{PalaciosCesnik2023,
  author    = {Palacios, R. and Cesnik, C.E.S.},
  title     = {Dynamics of Flexible Aircraft: Coupled Flight Mechanics, Aeroelasticity, and Control},
  publisher = {Cambridge University Press},
  series    = {Cambridge Aerospace Series},
  number    = {52},
  year      = {2023},
  doi       = {10.1017/9781108354868}
}

\end{document}